\documentclass{article}

\usepackage[english]{babel}

\usepackage[a4paper,top=2cm,bottom=2cm,left=3cm,right=3cm,marginparwidth=1.75cm]{geometry}

\usepackage{amsmath}
\usepackage{graphicx}
\usepackage[colorlinks=true, allcolors=blue]{hyperref}
\usepackage{booktabs} 
\usepackage{pifont}
\usepackage{array}     
\usepackage{booktabs} 
\usepackage{siunitx}  
\usepackage{float}

\title{Strategies to Measure Energy Consumption Using RAPL During Workflow Execution on Commodity Clusters}
\author{Philipp Thamm, Ulf Leser}

\begin{document}
\maketitle

\begin{abstract}
In science, problems in many fields can be solved by processing datasets using a series of computationally expensive algorithms, sometimes referred to as workflows. Traditionally, the configurations of these workflows are optimized to achieve a short runtime for the given task and dataset on a given (often distributed) infrastructure. However, recently more attention has been drawn to energy-efficient computing, due to the negative impact of energy-inefficient computing on the environment and energy costs. To be able to assess the energy-efficiency of a given workflow configuration, reliable and accurate methods to measure the energy consumption of a system are required. One approach is the usage of built-in hardware energy counters, such as Intel RAPL. Unfortunately, effectively using RAPL for energy measurement within a workflow on a managed cluster with the typical deep software infrastructure stack can be difficult, for instance because of limited privileges and the need for communication between nodes. In this paper, we describe three ways to implement RAPL energy measurement on a Kubernetes cluster while executing scientific workflows utilizing the Nextflow workflow engine, and one additional method using IPMI. We compare them by utilizing a set of eight criteria that should be fulfilled for accurate measurement, such as the ability to react to workflow faults, portability, and added overhead. We highlight advantages and drawbacks of each method and discuss challenges and pitfalls, as well as ways to avoid them. We also empirically evaluate all methods, and find that approaches using a shell script and a Nextflow plugin are both effective and easy to implement for workflow users. Additionally, we find that measuring the energy consumption of a single task is straight forward when only one task runs at a time, but concurrent task executions on the same node require approximating per-task energy usage using metrics such as CPU utilization.
\end{abstract}

\section{Introduction}\label{Introduction}
During scientific work, a series of computational steps is often required to extract the desired information from a given dataset. Such pipelines of multiple independent programs for data analysis are commonly referred to as scientific workflows \cite{dayal_scientific_2009}. Scientific workflows are used in many areas, such as genomics \cite{fellows_yates_reproducible_2021} or remote sensing \cite{sudmanns_assessing_2020}. Most of the time, workflows are being optimized for fast execution speeds in order to reduce the time needed for data analysis. However, the increasing awareness of the growing greenhouse gas emissions of the Information and Communication Technology (ICT) sector, which already accounted for up to 2.8\% of all global emissions in 2022 \cite{freitag_real_2021}, emphasizes the need for energy-based workflow optimization.

To be able to optimize a workflow for energy-efficient execution, developers need access to information regarding the current energy consumption of their workflow. One way to achieve that is the utilization of built-in Intel RAPL energy counters \cite{phung_modeling_2018, khan_rapl_2018, david_rapl_2010}. While accessing these energy counters is relatively straightforward on local hardware through available software such as perf\footnote{https://perfwiki.github.io/main/, last accessed: \date{May 14, 2025}} or IPMI\footnote{https://www.intel.de/content/www/de/de/products/docs/servers/ipmi/ipmi-home.html, last accessed: \date{January 10, 2025}}, it can pose a significant challenge in more complex environments. With the help of these energy counters, the energy consumption of CPU, integrated graphics, I/O-controllers, cache and RAM can be tracked depending on the CPU model \cite{khan_rapl_2018}, leading to estimations of the energy consumption which can be utilized to predict the energy consumption of the whole system \cite{kavanagh_rapid_2019}.

In this paper, we present our lessons-learned from implementing multiple methods to read the RAPL energy counters while executing scientific workflows on a shared, managed commodity cluster using the Nextflow workflow management engine \cite{spisakova_nextflow_2023} and Kubernetes \cite{senjab_survey_2023}. We discuss the challenges of reading RAPL values posed by the underlying infrastructure and present several solutions, considering a setup that leverages Nextflow and Kubernetes. Additionally, we provide an overview of the technologies employed, as well as a perspective on potential future work to facilitate the use of the information provided by RAPL.

In the following, we provide some background about the used technologies in Section \ref{Background}. Then, we outline the general circumstances leading to problems with the automated measurement of energy consumption via RAPL in our setup in Section \ref{Problems} before presenting multiple ways for automated energy measurement despite these pitfalls in Section \ref{Measurement_Strategies}. Our experimental results are presented in Section \ref{Experiments}. This is followed by a discussion of our findings in Section \ref{Discussion}. Section \ref{Future_Work} highlights opportunities for future work. Finally, we conclude our work in Section \ref{Conclusion}.

\section{Background}\label{Background}
This section discusses the most important technologies related to scientific workflows (\ref{Scientific_Workflows}), energy measurement (\ref{Methods_of_Measurement}) and especially RAPL (\ref{RAPL}), as well as workflow infrastructure (\ref{Cluster_Computing}).
In the context of scientific workflows on compute clusters, several key terms must be clearly distinguished. For this work, a cluster is a set of independent machines working together to execute computations efficiently. Each machine in the cluster is called a node, which is an independent computer capable of executing tasks. Nodes contain at least one CPU, which itself consists of multiple cores that can handle separate workloads simultaneously. A task is a distinct executable computational step within a workflow, defined by its inputs and outputs, and can be reused across workflows.

\subsection{Scientific Workflows}\label{Scientific_Workflows}
The computational analysis of large amounts of scientific data is often complex, leading to significant computational requirements. Many analysis pipelines consist of multiple interdependent steps, known as tasks. These pipelines of linked, interdependent data analysis tasks are called scientific workflows \cite{gil_examining_2007}. Due to the large inputs, high computational requirements and the segmentation into interdependent tasks, which can often be executed in parallel, scientific workflows are often executed on compute clusters consisting of multiple compute nodes (Section \ref{Cluster_Computing}).

When composing a scientific workflow, researchers define the inputs, workflow steps, and their dependencies. They use workflow languages to express this information in a format that the executing machine can interpret. These workflow languages are often embedded into workflow management systems, programs designed to facilitate the composition and implementation of workflows \cite{dayal_scientific_2009}. Additionally, they also help to make workflows more reproducible. Workflow management systems have varying feature sets. For instance, Galaxy \cite{the_galaxy_community_galaxy_2022} provides a graphical user interface (GUI) for the implementation and execution of workflows, improving its accessibility. Other workflow management systems like Snakemake \cite{molder_sustainable_2021} or Nextflow \cite{di_tommaso_nextflow_2017} use domain specific languages (DSL) to enable workflow implementation. Due to utilizing a DSL, the code written in these workflow management systems might be easier to write and read compared to general programming languages, since the features of the language are specifically tailored to scientific workflows. The use of a textual workflow language also improves portability and enables easy versioning through version management tools like Git.

\subsubsection*{Nextflow}
Nextflow \cite{di_tommaso_nextflow_2017} is a popular workflow management system primarily used in the area of Bioinformatics. Nextflow workflows are written in Groovy, a DSL based on Java. It offers support for containerization and orchestration technologies such as Docker (Section \ref{Docker}) and Kubernetes (Section \ref{Kubernetes}), among others. For this reason and due to the availability of multiple different workflows from the area of Bioinformatics, we chose to use Nextflow for our experiments.

\subsubsection*{Optimizing Workflow Execution}
Due to the complex structure of scientific workflows consisting of multiple interdependent tasks, optimization to a specific infrastructure is not trivial. Possible goals of optimization are minimal makespan \cite{bader_lotaru_2024} (the total execution time of the workflow from beginning to end), or minimal memory utilization for each individual task \cite{lehmann_ponder_2024}. Sometimes, approaches aim for multi-objective optimization \cite{khaleghzadeh_bi-objective_2021}. Other situations require in-budget optimization, where a workflow is optimized while adhering to specific constraints \cite{li_cost_2018}. Another possible objective is the reduction of a scientific workflows energy consumption \cite{durillo_moheft_2012}. To minimize the energy consumption of scientific workflows and individual workflow tasks, researchers need access to data regarding the energy consumption of the current configuration. Therefore, researchers require reliable methods to measure the task energy consumption of a workflow.

\subsection{Methods of Energy Measurement}\label{Methods_of_Measurement}

\subsubsection*{Physical Measurement}
The first method to measure the power consumption of a device is through the use of an external power meter. These measurement devices are generally positioned between the wall socket and the power supply of a device or computing node. They measure the total power consumption of the entire computing node. Since the entire power flowing to the node also flows through the power meter, these devices are typically very accurate \cite{jay_experimental_2023}. They also have the advantage that their impact on the measured system is negligible. On the other hand, installing power meters requires an additional financial investment, and it is not possible to individually measure the power consumption of concurrently running tasks or specific hardware components such as CPU, DRAM, disk or power supply. Additionally, this approach to measuring power is impractical for large systems, since an additional power meter is required for each node.

\begin{figure}
\centering
\includegraphics[width=0.5\linewidth, trim={0cm 0cm 0cm 0cm},clip]{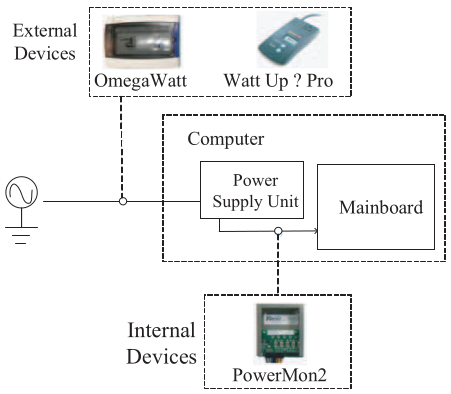}
\caption{The installation locations for external and internal physical power meters \cite{lin_taxonomy_2021}.}
\label{fig:Power_Meters}
\end{figure}

Alternatively, intra-node power meters can also be used for physical measurements. Unlike external power meters, these devices are placed \textit{inside} of a computing node. They can be placed between the power supply and the main board, such as PowerMon2 \cite{bedard_powermon_nodate}, or next to specific components, such as PowerInsight \cite{laros_powerinsight_2013}. Intra-node power meters can measure the power consumption of specific components such as CPU, GPU or cooling fans, but they share many drawbacks with external power meters. In addition, they are often expensive to install and inflict additional liability issues. 
Figure \ref{fig:Power_Meters} shows typical installation locations of external and internal physical power meters on a computer.

\subsubsection*{Approximation through Energy Models}
 Energy models use a set of metrics collected during or after the program execution to calculate an approximation of the energy consumption. These metrics can include CPU frequencies, run time, the thermal design power of the processor, the number of cache misses, fan speed and many more. Based on these metrics, numerous formulas to calculate task power consumption have been proposed \cite{lin_taxonomy_2021}. An advantage of energy models is that they do not require any additional hardware or the existence of specific feature sets like hardware energy counters. The most important disadvantage of software-based energy models is that it is hard to define a metric that works well across different machines, architectures and software \cite{obrien_survey_2018}. One model that performs well for one task on a specific machine might perform much worse when used in a different context, making it much less useful when applied outside the context it was originally optimized for.

\subsubsection*{Hardware Energy Counters}
Most component manufacturers embed digital sensors or onboard measurement circuits to measure the power consumption of the entire system, the CPU cores, the memory or other components in their products.

Intel introduced the Running Average Power Limit (RAPL) in 2011 \cite{david_rapl_2010} in the Sandy Bridge architecture as a software power model based on architectural events including the cores, integrated graphics and I/O. In the Haswell architecture, this implementation was exchanged with a version based on fully integrated voltage regulators, enabling actual power measurement and improving the accuracy of RAPL \cite{hackenberg_energy_2015}. More information about Intel RAPL is presented in Section \ref{RAPL}.

AMD introduced a version of hardware energy counters similar to the first version of RAPL with their Zen architecture. Similar to the early version of RAPL, this implementation can provide inconsistent results due to using execution path-based modeling instead of on-chip energy measurement \cite{schone_energy_2021}.

NVIDIA provides users with an API called NVIDIA Management Library (NVML) that provides GPU device metrics such as current utilization, temperature and power draw \cite{sen_quality_2018}.

The advantage of hardware energy counters is the typically good accuracy compared to purely software based models due to the deep integration in the system components and the utilization of integrated measurement hardware \cite{hackenberg_energy_2015, schone_energy_2021}. One drawback is that only the system components equipped with appropriate measurement hardware by the hardware designer can be measured. For example, RAM and the on-chip I/O-controllers are measurable in some implementations of Intel RAPL \cite{khan_rapl_2018}, but other components such as secondary storage (e.g., disks or solid state drives) and interconnecting network are not equipped with appropriate energy counters, and therefore can not be measured. Additionally, available interfaces and domains for energy measurement based on hardware energy counters are vendor-specific and can vary between models. This limits the portability of software utilizing hardware energy counters.

\subsection{Running Average Power Limit}\label{RAPL}
Running Average Power Limit (RAPL) from Intel is an interface to estimate power usage that is built into many Intel CPUs \cite{david_rapl_2010, hackenberg_energy_2015}. It was first implemented as a software power model in 2011 \cite{david_rapl_2010} in the Sandy Bridge architecture. Starting with the Haswell architecture, it was changed to using fully integrated voltage regulators, making actual power measurement instead of estimations possible and thereby improving the accuracy of the results \cite{hackenberg_energy_2015}.

\begin{figure}
\centering
\includegraphics[width=0.5\linewidth, trim={0cm 0cm 7cm 3.21cm},clip]{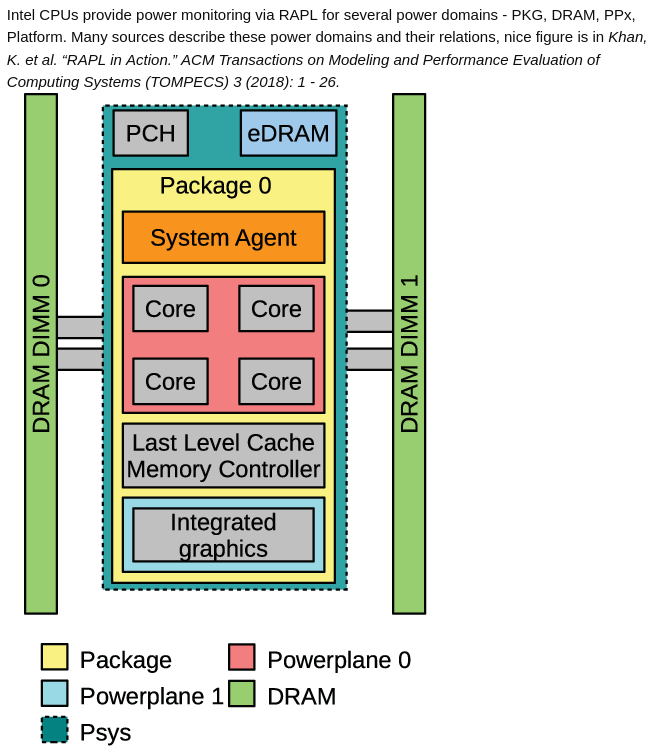}
\caption{An overview of the system components covered by each RAPL domain \cite{khan_rapl_2018}.}
\label{fig:RAPL}
\end{figure}

RAPL is capable of reporting the energy consumption in different power domains. Figure \ref{fig:RAPL} shows an overview of the supported domains and which parts of the system are contained in each of them.

\begin{itemize}
\item The Core domain (here labeled Powerplane 0) reports the energy consumption of all CPU cores added together. Note that RAPL is not capable of reporting the power usage of an individual CPU core.

\item The Graphics domain (Powerplane 1) reports the energy usage of the integrated graphics component. If the chip has an eDRAM component (a small amount of RAM embedded directly on the chip, acting akin to a Layer 4 cache), its power consumption is also included in the Graphics domain. 

\item The Package domain contains both the CPU cores and the integrated graphics, in addition to some  components not contained in the Core or Graphics domain. This includes higher level caches which are not directly associated with a CPU core, the memory controller, system agent and the I/O controller (not shown).

\item The DRAM domain is separate and reports the power usage of all the DRAM modules added together. It is important to note that the power used by the DRAM is not included in any other domain. This makes the DRAM domain completely separate of all the other domains, and the only domain that logs exclusively the power consumption of components not located on the CPU chip.

\item The Psys domain reports the power consumption of almost the whole system. It includes the Package domain and most additional components on the CPU, as well as some additional system components such as cooling fans. Notable exceptions are the DRAM, which is exclusively monitored by the DRAM domain, and secondary storage like disks, which is not monitored by RAPL.
\end{itemize}

Not all RAPL domains are available on all CPUs, but the availability of each domain depends on the microarchitecture and individual model of CPU. Documentation about each model of CPUs is available in the respective specifications on Intel's website\footnote{https://www.intel.com/content/www/us/en/ark/products/series/122139/intel-core-processors.html, last accessed: \date{May 14, 2025}}.

RAPL energy consumption reports can be accessed through Model-specific registers (MSR). The values in these registers are expressed in energy units and represent the energy consumed in microjoules ($\mu J$) since the processor was started, multiplied by a static conversion value that depends on the domain and the individual model of CPU. Each MSR that contains the energy counter for one of the domains is updated roughly every 1 ms (1000 Hz) with low overhead \cite{khan_rapl_2018}. The RAPL power monitoring is always running, starting when the machine is booted up. The machine-specific registers (MSRs) are limited in their number of bits. Some registers are limited to 32 bits, others to slightly larger values, e.g., 36 bits, depending on the CPU model. Therefore, each energy counter occasionally overflows and loops around to zero. This is not logged or communicated in any way. For example, an Intel(R) Xeon(R) Silver 4314 CPU completed a loop from zero to overflowing in 52 minutes while executing a workload of mixed intensity during our experiments. Therefore, measuring the energy over a specific time frame requires continuous monitoring of the MSR to catch overflows and factor them into the calculated energy value. Additionally, timestamps are not automatically attached to the read values.

In terms of general accuracy of energy measurement conducted using Intel RAPL, earlier research concludes that the energy consumption reported by RAPL is reliable for both CPU \cite{jay_experimental_2023} and memory \cite{alt_experimental_2024}. Currently, we can not confirm if this is also the case for scientific workflows executed on compute clusters, since we do not have access to physical power meters for accurate power readings on the cluster hardware.

\subsection{Workflow Infrastructure}\label{Cluster_Computing}
Clusters are distributed systems \cite{van_steen_brief_2016} that allow for fast execution of tasks by providing computational resources beyond what a singular machine can offer. A cluster consists of a number of independent computing elements, called nodes. Each node is a complete system with at least its own CPU and memory. These nodes are interconnected with each other to enable communication and data transfer. Usually, one of the nodes in the cluster functions as the management node, controlling and scheduling the worker nodes. The user communicates with the cluster through the management node, which makes the whole cluster appear as a singular structure to the user. To make this form of interaction of the user with the cluster possible, an orchestration system is needed to schedule available resources of the cluster transparently and automatically. Kubernetes \cite{carrion_kubernetes_2023} is a widely used container orchestration system that is often used to coordinate the execution of workloads, including scientific workflows, on compute clusters.

\subsubsection*{Kubernetes}\label{Kubernetes}
Kubernetes \cite{carrion_kubernetes_2023} is a container orchestration system that was originally created by Google. It was donated to the Cloud Native Computing Foundation in 2016 and is now open source. In Kubernetes-managed environments, a pod represents an execution unit that is assigned to a node with a defined amount of resources (CPU, memory). Typically, each pod executes a single task, except for a dedicated command-and-control pod, which orchestrates workflow execution. Each pod encapsulates a container, which contains the program along with all its dependencies, ensuring a consistent execution environment for a task. Kubernetes can be used to deploy, monitor and scale containers on a compute cluster or other distributed architectures. It implements the automatic scheduling and re-scheduling of containers on a distributed infrastructure, and also supports automatic resource allocation and load balancing. These features make Kubernetes very useful when workflows are executed on a compute cluster. Kubernetes can work with multiple types of containers. These containers are often managed by Docker \cite{rad_introduction_2017}, a platform for sharing and managing containers. 


\subsubsection*{Docker}\label{Docker}
The containers utilized by Kubernetes are built and managed by Docker \cite{rad_introduction_2017}. Docker is an open platform to package, distribute and deploy applications in a secure, portable and lightweight manner. It allows users to package applications into Docker images, together with all their needed resources and environments. These Docker images can then be shared and run on other machines with no additional configuration through the Docker Engine. This process is much more lightweight than virtualization \cite{rad_introduction_2017}. At the same time, Docker images run faster and similarly secure since each container runs in its own virtual environment with no access to resources outside of it. These characteristics make Docker containers ideal for executing the tasks of a workflow. They allow the user to deploy different applications used in the same workflow independently on any node in the cluster, sequentially or in parallel. At the same time, they are secure and no additional configuration on the nodes is necessary. 


\section{Challenges for Automatic Measurement of Energy Consumption of Scientific Workflows Executed on Compute Clusters}\label{Problems}
To determine the energy consumption of a machine over a timeframe, one can use RAPL energy counters. The counter values can be read from the respective MSRs and written to a text file. Time stamps are manually added when writing new energy counter values to the file in order to make correlation of the values with a specific point in time possible. This way, it is possible to calculate the energy consumption of the machine over any period of time by calculating the difference between the values stored in the energy counters at the starting point and at the end of the time period.

In order to measure the energy consumption of a scientific workflow, the RAPL values are collected right before the workflow is started and right after the workflow has finished. By subtracting the value of the energy counter before execution from the value of the energy counter after execution, the amount of energy consumed by executing the workflow can be calculated, provided that the workflow was the only workload executed on the node during measurement. Task-wise energy measurement is also possible by the same method (reading RAPL energy counters at the start of the task and when the task is finished), but again only if the task is the only workload executed on the machine. If multiple workloads are executed on the same machine in parallel (e.g., multiple tasks from the same scientific workflow or from two workflows running concurrently), an additional heuristic such as the CPU time used by each of the tasks must be applied, because RAPL only captures the total energy consumption over the time period.

The fact that the workflows are run on a cluster consisting of multiple independent machines orchestrated by Kubernetes leads to some inherent complications for power measurement using RAPL when compared to a local execution. In this section, the issues arising and possible steps to be taken are discussed.

\subsection{Multiple Nodes}\label{Multiple_Nodes}
The workflows are executed on a cluster consisting of multiple independent nodes with their own hardware, including the CPU. It is not clear in advance which and how many nodes of the cluster will be used during the execution of the workflow. The user only has a limited influence on the scheduling in Kubernetes by specifying parameters such as the utilized namespace or the number of splits of a data set, leading to more or less parallelism. Therefore, the energy consumption of each node that could \textit{potentially} participate in the execution of \textit{any} part of the workflow needs to be measurable. However, RAPL is only capable of measuring the power consumption on the same local chip, but not remotely for other CPUs. That means that the power consumption of each CPU in each node that could potentially be used during the workflow execution must be prepared for measurement. If a workload (i.e., one of the tasks) belonging to the measured scientific workflow is executed on any of the machines, its energy consumption must be measured over that time period. After workflow execution has finished, the power consumption of all the involved nodes has to be added to calculate the total energy consumption.

\subsection{Required Privileges}\label{Root_Privileges}
The program measuring energy consumption requires root privileges to access the values stored in the RAPL registers. This is necessary because the fine-grained energy measurements provide a vector for side-channel attacks by closely monitoring the consumed energy and reconstructing the executed instructions from these measurements. It has been proven that such attacks are feasible \cite{lipp_platypus_2021}. For instance,  DeepTheft \cite{gao_deeptheft_2023} is a tool designed to steal Deep Neural Network weights through an Intel RAPL side-channel attack.

For this reason, one can not access the RAPL registers from the typical container process running on the cluster. Instead, a container with the required privileges to read RAPL registers must be deployed on each utilized node.

\subsection{Constrained Docker Containers}\label{Isolated_Docker}
In a Kubernetes setup, each task of the workflow, including the main control task, are running in isolated Docker containers, which means that the tools at their disposal are limited. The typically very small Docker containers often lack even basic system tools, simply because they are not needed for the task performed in the container. For example, the vi text editor, normally included with any Unix-based operating system, is sometimes missing in small Docker containers. This reduces the amount of options available to control the process of RAPL measurement without deploying custom containers with added tools.

\subsection{Requirement of Continuous Measurement}\label{Continuous_Measurement}
Another factor that complicates the process of obtaining accurate energy values from RAPL energy counters is that a single read-out before and after each task of the workflow is not sufficient. The used power is represented in the RAPL register as energy units in microjoule ($\mu J$) \cite{khan_rapl_2018}. Due to the small unit being utilized and the register being limited to 38 bits, the number in the register overflows to zero between every half an hour up to every 24 hours, depending on the computational requirements of the workload. These overflows are not logged or indicated in any way. To avoid reporting wrong values due to an undetected overflow, it is necessary to constantly monitor the RAPL register in sufficiently small time intervals to detect overflows, so they can be considered when calculating the total amount of used power.

\subsection{Available RAPL Domains}\label{Available_Domains}
\begin{figure}
\centering
\includegraphics[width=0.9\linewidth, trim={0cm 0cm 0cm 0cm},clip]{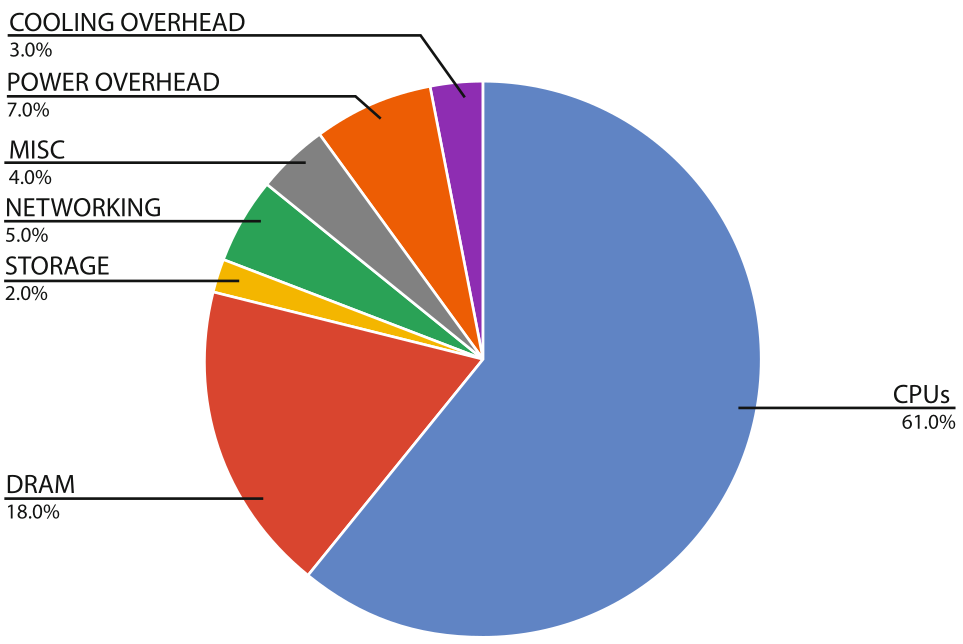}
\caption{A breakdown of the typical power consumption of a physical server \cite{barroso_datacenter_2018}. The figure assumes two-socket x86 servers and 12 DIMMs
per server, and an average utilization of 80\%.}
\label{fig:Power_Consumption}
\end{figure}

Not all RAPL domains presented in Section \ref{RAPL} are available on every CPU. For instance, the Intel(R) Xeon(R) Silver 4314 CPUs running in the cluster used for this research support only the package and DRAM domains. Fortunately, these two RAPL domains contain almost all the components that are included in any of the other RAPL domains, providing an almost complete picture of the energy consumed by the components monitored by RAPL. These components account for about 63\% of the total energy consumed by a typical server \cite{lin_taxonomy_2021}, and up to 79\% if the average utilization is high \cite{barroso_datacenter_2018}. The other 21\% are consumed by the disk (2\%), the network (5\%) and other parts of the system (14\%) such as cooling fans or the power supply. Figure \ref{fig:Power_Consumption} shows a breakdown of the typical power consumption of a server with an average utilization of 80\%.

\subsection{Granularity}\label{Granularity}
To be able to use the data collected during energy measurement for workflow optimization, it is important to measure with high granularity. Here, one needs to differentiate between workflow granularity and hardware granularity. In terms of workflow granularity, it would be ideal to measure the energy consumption of the whole workflow, of each physical or abstract workflow task, and of each individual container that is part of the workflow. From the hardware perspective, the goal is to measure the energy consumption of each node, each individual CPU core on a node, and each individual thread running on a core. In practice, the limiting factor is the availability of RAPL domains. If the only available domain is the Package domain, it is not possible to measure individual cores or threads, but only the energy consumption of the whole CPU for each individual CPU in the system. From the workflow perspective, this means that the energy consumption of a workflow, task or container can only be individually measured if no other software is running on the same machine at the time of execution. If some other software is running at the same time as the workflow, task or container, the energy consumption can only be estimated by subtracting the energy consumed by that other software utilizing a separate metric.

\section{Measurement Strategies}\label{Measurement_Strategies}
In this section, we present four measurement strategies we developed in order to achieve reliable and accurate energy measurements for workflow executions in Nextflow on a Kubernetes cluster utilizing Intel RAPL and IPMI. We explain each strategy in detail and discuss their respective advantages and drawbacks.

\subsection{Design Goals for Energy Measurement}\label{Design_Goals}
We identified a set of eight characteristics an ideal approach for RAPL-based energy measurement of workflow tasks should have.

\subsubsection*{No additional software necessary}
An ideal approach for RAPL-based energy measurement works without additional software that needs to be installed and maintained separately on the cluster by the cluster-administrator or the workflow-administrator. Depending on additional software makes an approach for measurement harder to configure and limits the scope where that method can be applied, since the needed software might not be available when working with different clusters or workflow systems.

\subsubsection*{Self-contained on the cluster}
Approaches for energy measurement should not require a continuous connection to an external device from the cluster, such as the user's local machine. If such a connection is necessary, it limits how the user can utilize their machine while the workflow is running on the cluster. Most importantly, the local machine would need to stay active and connected to the network in order to ensure that continuous communication between the local machine and the cluster is possible. This is a significant disadvantage in usability, especially for long-running workflows where the user might want to turn off their local machine or leave the network, like for executions of workflows overnight.

\subsubsection*{Dealing with workflow faults}
Software for energy measurement should be able to react to faults or unexpected changes in the number and order of executed workflow tasks, and still be able to measure the energy consumption of each task correctly. Otherwise, unexpected events during workflow execution might lead to incomplete or wrong energy consumption data.

\subsubsection*{Full measurement for all workflow tasks}
An ideal approach for RAPL-based energy measurement should be capable of capturing the full energy consumption individually for all physical tasks in a workflow. The measurements for the physical tasks can then be aggregated to represent the energy consumption of logical tasks or the whole workflow. If the energy data for some tasks is incomplete, its usefulness for future usage will be limited.

\subsubsection*{Easy portability to other workflows and workflow systems}
Implemented approaches for energy measurement should be workflow-agnostic and easy to implement for a new workflow. Ideally, they are also portable to other workflow systems to enable broad usage.

\subsubsection*{No workflow modifications necessary}
Enabling energy measurement for a workflow should require no changes to the structure or implementation of the workflow. Such requirements would make implementation of compatible workflows more tedious and difficult, and therefore limit the usability of the tools for measurement.

\subsubsection*{Low overhead}
Every approach for RAPL-based energy measurement will add \textit{some} computational overhead during workflow execution. This computational overhead will cause additional load on the system and increase the energy consumption. It is important that the extent of this overhead is kept small and limited to the time of workflow execution.

\subsubsection*{Multi-tenancy}
Methods for RAPL-based energy measurement should be capable of measuring the energy consumption of multiple independent scientific workflows running in parallel on the same nodes of the cluster. 

\subsection{Approaches}\label{Approaches}
In the following, four approaches to monitor workflow energy consumption by reading values from RAPL energy counters are presented, focusing on Nextflow as workflow management system and Kubernetes as container orchestration system. Except for the method using the API of Prometheus, all methods use dedicated pods, running on each node of the cluster, that were configured with special permissions to be able to read from RAPL registers. When receiving a command, these pods execute a script to continuously read the RAPL energy counters and store the values in log-files together with time stamps. These log files are stored on the shared storage of the cluster and can be accessed by any pod. As soon as a physical task is finished, its energy consumption can be calculated either directly on the cluster or on the local machine of the workflow user by copying the file. When the workflow is completed, its energy consumption can be calculated by summing up the values for all physical tasks. When using Prometheus, the energy consumption does not have to be calculated manually. Prometheus automatically converts the measured values into consumed energy. The consumed energy for a specific time interval can be directly requested using the API. Within this general setting, the four methods differ in how they coordinate the process of measurement. Since reading and logging RAPL values continuously would waste valuable system resources, including CPU overhead and required storage, it is more efficient to only measure RAPL values when they are needed. The proposed methods offer different strategies to measure RAPL counters only during workflow execution, or otherwise circumvent the resource overhead introduced by continuous measurement.

Table \ref{tab:comparison} shows which of the eight design goals for energy measurement are fulfilled by each of the four approaches. A checkmark ("\ding{51}") means that the criterion is fulfilled completely, while the cross ("\ding{55}") denotes a criterion that is not fulfilled. A checkmark in brackets ("(\ding{51})") symbolizes a criterion that is technically not completely fulfilled. For example, the Plugin method technically requires the plugin as additional software. However, if the plugin was officially released, it can be integrated into a Nextflow workflow using one line of code or one command line argument. It is automatically downloaded and installed during workflow execution and therefore does not cause any additional work for the workflow developer or the workflow user. For this reason, the criterion can be considered fulfilled.

\begin{table}[h]
    \centering
    \caption{Comparison of Energy Measurement Methods}
    \label{tab:comparison}
    \resizebox{\textwidth}{!}{ 
    \begin{tabular}{lcccc}
        \toprule
        \textbf{Feature} & \textbf{Part of Workflow} & \textbf{Shell-Script} & \textbf{Plugin} & \textbf{Prometheus} \\
        \midrule
        No additional software            & \ding{51}  & \ding{51}  & (\ding{51})  & \ding{55}  \\
        Self-contained                & \ding{51}  & \ding{55}  & \ding{51}  & \ding{51}  \\
        Dealing with workflow faults                 & \ding{55}  & \ding{51}  & \ding{51}  & \ding{51}  \\
        Enables full measurement            & \ding{55}  & \ding{51}  & (\ding{51})  & \ding{51}  \\
        Easy portability                             & \ding{55}  & \ding{51}  & \ding{51}  & \ding{51}  \\
        No workflow modifications        & \ding{55}  & \ding{51}  & (\ding{51})  & \ding{51}  \\
        Low overhead & \ding{51}  & \ding{51}  & \ding{51}  & \ding{55}  \\
        Multi-tenancy & \ding{55}  & \ding{55}  & \ding{55}  & \ding{55}  \\
        \bottomrule
    \end{tabular}
    }
\end{table}

\subsubsection{Manage Energy Measurement as part of the Workflow}\label{Inside_Workflow}
\begin{figure}
\centering
\includegraphics[width=0.9\linewidth, trim={0cm 0cm 0cm 0cm},clip]{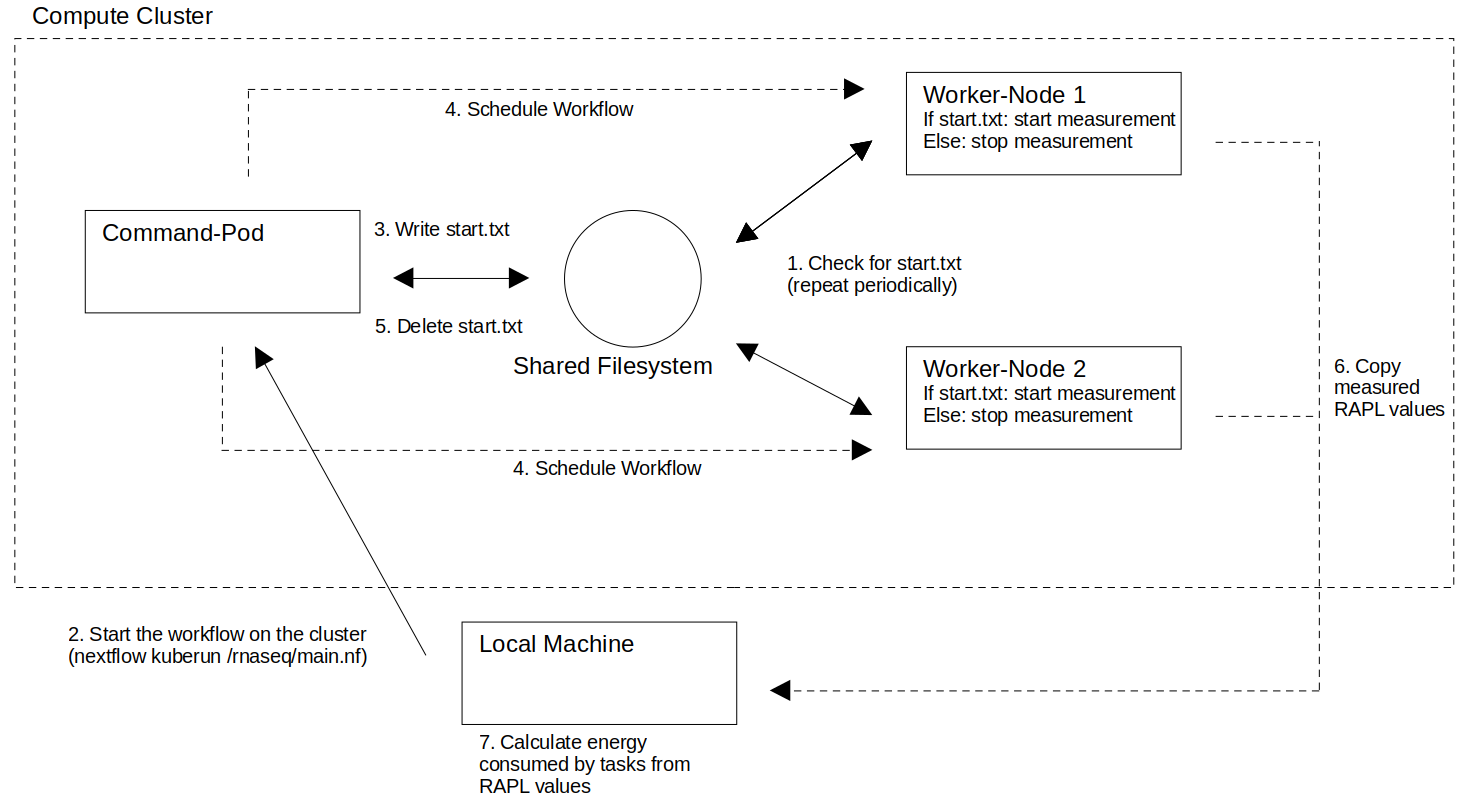}
\caption{The communication between the machines when controlling the RAPL measurement as part of the workflow. Each arrow represents a communication between two components of the cluster, or between a component of the cluster and the local machine of the user running the workflow. Solid arrows represent communication directly related to the process of measuring RAPL values, while dashed arrows show communication for workflow execution and extraction of results. The numerated annotations describe the actions that are part of the measurement process in ascending order, starting with 1. as the first action and ending with 7. as the last action of the measurement process. Note that the figure shows two worker nodes and one separate command pod. In practice, the number of used worker nodes depends on the cluster configuration and on the executed workflow. The command pod can be executed on a separate node of the cluster or on one of the worker nodes. During the process of energy measurement, it can be treated like any other pod that is part of the executed workflow.}
\label{fig:Inside_Workflow}
\end{figure}

In this strategy, the measurement of the RAPL values is managed directly as part of the workflow. The code of the workflow is modified to include additional code that starts and stops the measurement in pods configured with root privileges. That means that in order to utilize this technique for task-based measurement on individual nodes of the cluster, the code of each individual task in the workflow needs to be changed accordingly. Since a direct communication between pods is not possible due to the limitations described in Section \ref{Isolated_Docker}, a workaround is necessary. The workflow is extended with two additional tasks which are executed at the start and at the end of the workflow. The first task writes a file start.txt to a specific location in the file system. The pods for energy measurement continuously check this location in the file system using a daemon that runs in the background. As soon as the file start.txt is detected, they begin the energy measurement by writing the values of the energy counters stored in the RAPL MSRs to a file together with time stamps. As long as the file exists, each pod for energy measurement continues to append the current values of the registers to the file in regular intervals. At the end of the workflow, the second additional task is executed to remove the file start.txt. When that happens, the energy measurement is stopped. The energy values are stored on the shared storage of the cluster, and can therefore be accessed by any pod. A separate script is then used to calculate the energy consumed by each physical task and copy the results to the local machine of the workflow user. This process enables communication between the pods through the file system and ensures that the energy is measured during the execution of the workflow, but not when the workflow is finished. In order to enable energy measurement for individual physical tasks, the same technique is used to write and delete the file used to start measurements before and after the respective task. Figure \ref{fig:Inside_Workflow} shows the communication between the machines during the workflow in order to enable energy measurement.

%

\subsubsection*{Advantages and Drawbacks}

The method implementing the energy measurement as part of the workflow does not fulfill four of our seven criteria of a good solution for RAPL-based energy measurement. An overview of all criteria for an ideal energy measurement strategy utilizing Intel RAPL and which of them are fulfilled by the individual presented approaches can be found in Table \ref{tab:comparison}. Table \ref{tab:strengths_weaknesses} shows a general overview of the strengths and weaknesses of each of the presented approaches.

\begin{itemize}
    \item \textbf{No additional software:} No additional software is required for energy measurement. All changes to the software are in the workflow itself.
    \item \textbf{Self-contained:} The approach does not rely on additional external hard- or software, making it self-contained.
    \item \textbf{Dealing with workflow faults:} The measurement process is unable to react to workflow faults. A failing workflow leads to the following tasks not being executed, which also means that the file used as a signal is not deleted automatically and the energy measurement continues. In that case, a manual deletion of the file is necessary in order to stop the energy measurement.
    \item \textbf{Enables full measurement:} The energy measurement is not complete because it is started and stopped by a task of the workflow. Therefore, the energy measurement can only be started when the workflow is already running and other tasks are already being scheduled, and stopped when it is not finished yet. That means at least the first and last task of the workflow can not be measured completely, leading to slightly incomplete measurements for some tasks of the workflow.
    \item \textbf{Easy portability:} The method is not easily portable to other workflows or even to measure only individual tasks, since additional tasks to write and remove start.txt have to be added to every workflow script individually. If the energy of a new workflow shall be measured, it is necessary to add the code for the measurement.
    \item \textbf{No workflow modifications:} Workflow modifications are necessary to utilize the method, since additional tasks are necessary to create and remove the file start.txt. This increases the workload for the user implementing and using the workflow.
    \item \textbf{Low overhead:} The overhead induced by this method is limited to the additional actions for writing to the file that controls the RAPL monitoring. This amount of additional overhead is negligible. Additionally, two additional tasks for controlling the energy measurement need to be scheduled. However, these tasks are small and executed quickly, making their overhead insignificant.
    \item \textbf{Multi-tenancy:} Since RAPL does not support energy measurements for individual CPU cores, an additional heuristic is necessary to assign a specific amount of energy to a physical task, if multiple tasks are running in parallel on the same CPU. The same is still true if the physical tasks belong to multiple individual workflows running in parallel.
\end{itemize}

\subsubsection{Wrapping Shell-Script}\label{Shell-Script}
\begin{figure}
\centering
\includegraphics[width=0.84\linewidth, trim={0cm 0cm 0cm 0cm},clip]{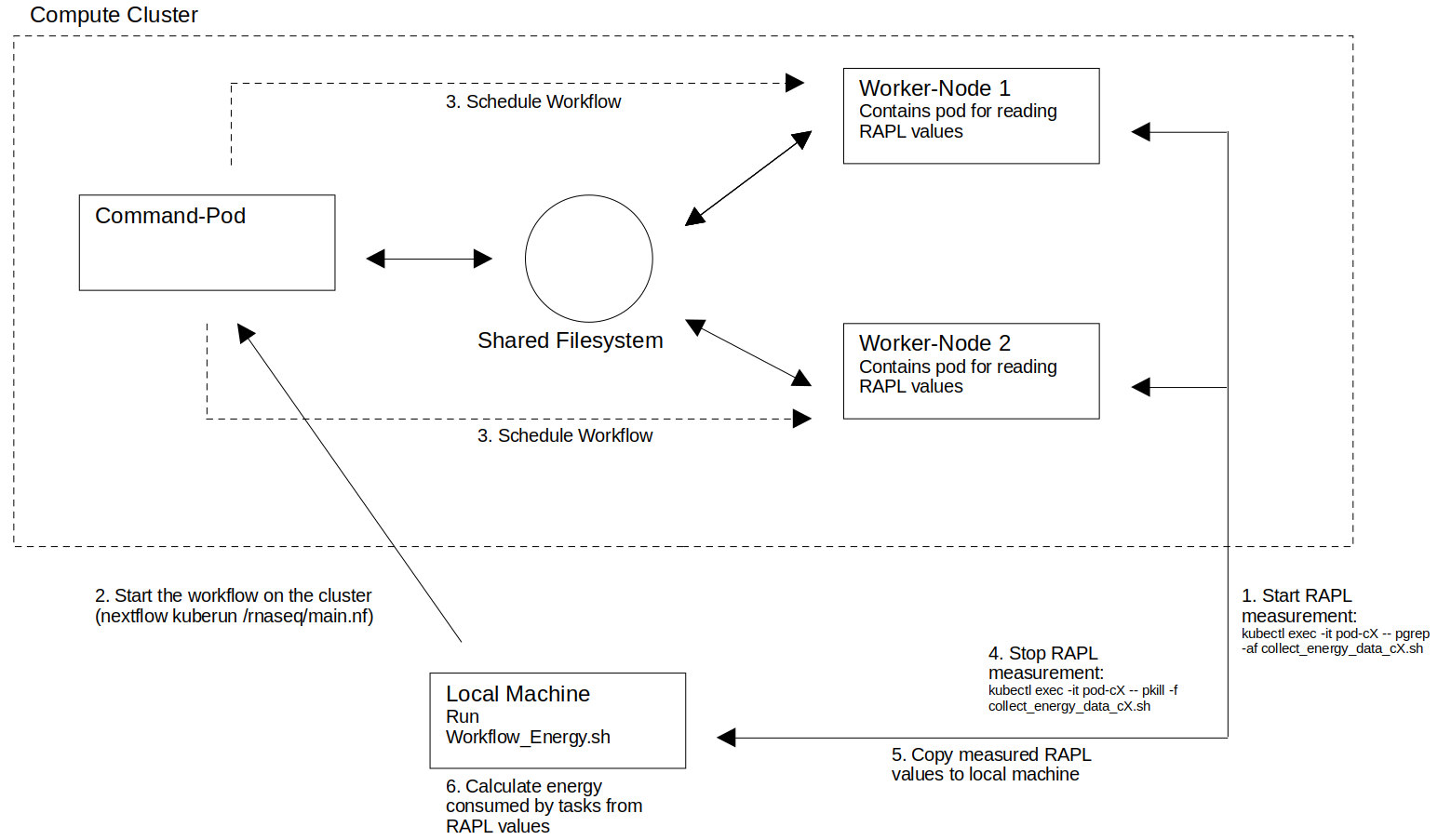}
\caption{The communication between the machines when controlling the RAPL measurement from the machine of the user using a shell-script. Each arrow represents a communication between two components of the cluster, or between a component of the cluster and the local machine of the user running the workflow. Solid arrows represent communication directly related to the process of measuring RAPL values, while dashed arrows show communication for workflow execution and extraction of results. The numerated annotations describe the actions that are part of the measurement process in ascending order, starting with 1. as the first action and ending with 6. as the last action of the measurement process. Note that the figure shows two worker nodes and one separate command pod. In practice, the number of used worker nodes depends on the cluster configuration and on the executed workflow. The command pod can be executed on a separate node of the cluster or on one of the worker nodes. During the process of energy measurement, it can be treated like any other pod that is part of the executed workflow.}
\label{fig:Shell-Script}
\end{figure}

The second method of measuring the energy of a scientific workflow does not change the workflow itself. Instead, the execution of the workflow is wrapped in a shell-script executed on the local machine of the user. This script automatically coordinates the workflow execution with the energy measurement. The shell-script first starts the energy measurement by sending commands to the pods for energy measurement using kubectl. Then the workflow is executed without any changes compared to a normal execution without energy measurement. The script periodically polls if the workflow is still running by checking the status of the coordinating pod of the workflow using kubectl commands. As soon as the workflow finishes running, the script stops the energy measurement and copies the files containing RAPL energy values with timestamps to the local machine of the user for further processing. Figure \ref{fig:Shell-Script} shows the individual steps of communication between the machines when using this method.

%
%

\subsubsection*{Advantages and Drawbacks}

\begin{itemize}
    \item \textbf{No additional software:} Except for the shell-script itself, no additional software is required on the cluster or on the local machine of the workflow user.
    \item \textbf{Self-contained:} A drawback of using a shell-script is that the energy measurement is not self-contained. The local machine of the user needs to run the script during the entire workflow. The machine needs to stay online and connected to the cluster in order to successfully orchestrate the energy measurement.
    \item \textbf{Dealing with workflow faults:} The shell-script can monitor the workflow and the state of the cluster during execution. It is therefore capable of reacting to workflow faults.
    \item \textbf{Enables full measurement:} Since the shell-script controls both the workflow execution and the measurement process, it can start the measurement before initiating the workflow execution. For this reason, full measurement of all workflow tasks is possible.
    \item \textbf{Easy portability:} The shell-script is independent of the executed workflow. To use the script for monitoring a different workflow, only the single line initiating workflow execution must be changed. 
    \item \textbf{No workflow modifications:} This method does not require any modifications of the workflow.
    \item \textbf{Low overhead:} This method does not introduce any overhead in the workflow itself. The overhead introduced by the shell-script while communicating with the cluster is negligible.
    \item \textbf{Multi-tenancy:} Since RAPL does not support energy measurements for individual CPU cores, an additional heuristic is necessary to assign a specific amount of energy to a physical task, if multiple tasks are running in parallel on the same CPU. The same is still true if the physical tasks belong to multiple individual workflows running in parallel.
\end{itemize}

\subsubsection{Integration through Nextflow-Plugin}\label{Nextflow_Extension}
\begin{figure}
\centering
\includegraphics[width=0.84\linewidth, trim={0cm 0cm 0cm 0cm},clip]{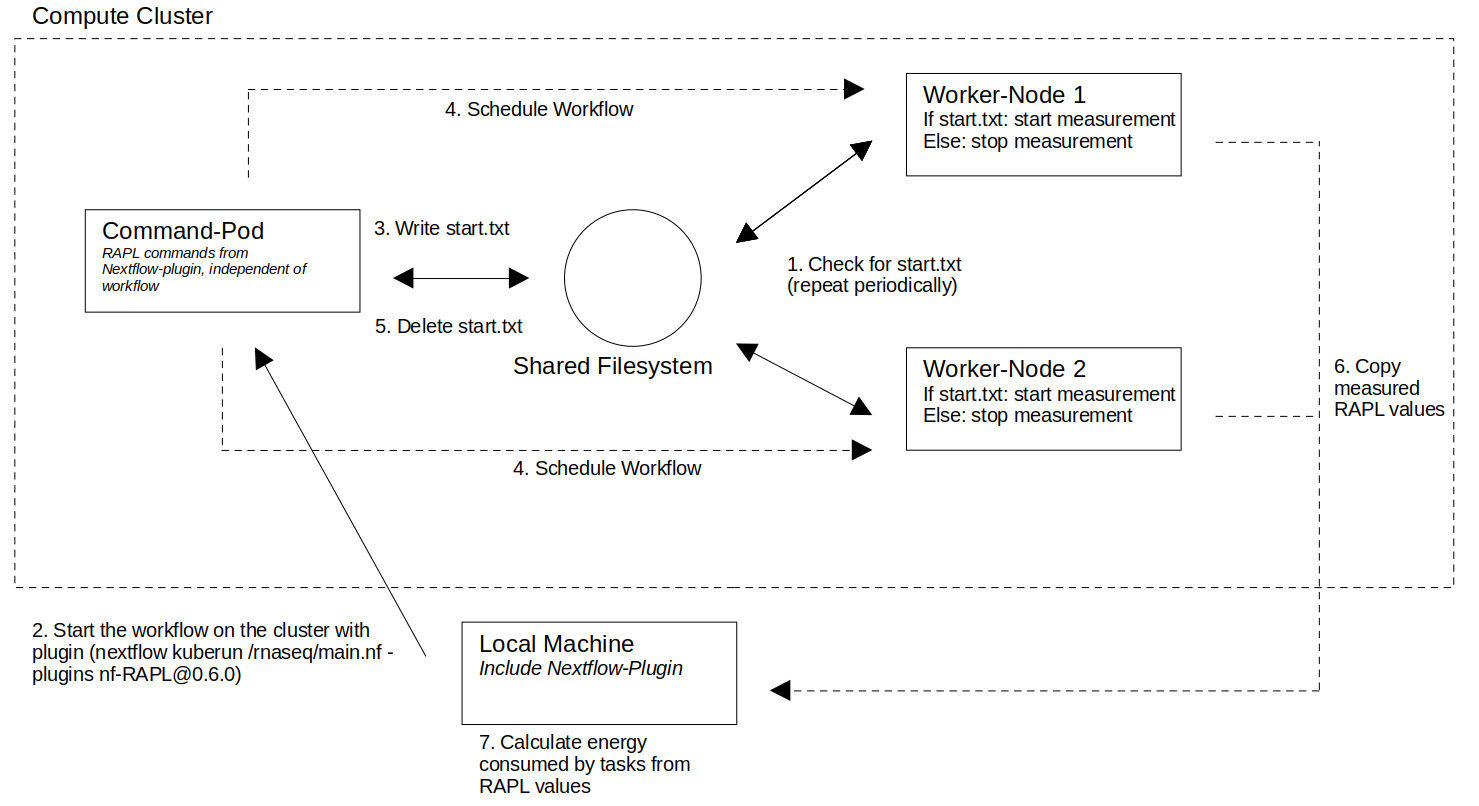}
\caption{The communication between the machines when controlling the RAPL measurement from inside Nextflow using a workflow-independent plugin that is executed in addition to the workflow. Each arrow represents a communication between two components of the cluster, or between a component of the cluster and the local machine of the user running the workflow. Solid arrows represent communication directly related to the process of measuring RAPL values, while dashed arrows show communication for workflow execution and extraction of results. The numerated annotations describe the actions that are part of the measurement process in ascending order, starting with 1. as the first action and ending with 7. as the last action of the measurement process. Note that the figure shows two worker nodes and one separate command pod. In practice, the number of used worker nodes depends on the cluster configuration and on the executed workflow. The command pod can be executed on a separate node of the cluster or on one of the worker nodes. During the process of energy measurement, it can be treated like any other pod that is part of the executed workflow.}
\label{fig:Extension}
\end{figure}

A third way of automating the energy measurement through RAPL is by using a Nextflow plugin. Nextflow provides support for plugins that are loaded and executed individually for any workflow. A Nextflow plugin is an extension that enhances Nextflow's functionality by adding custom features or integrations without modifying the core workflow definition. Plugins can hook into different stages of execution, allowing developers to automate specific tasks before, during, or after a workflow runs. A solution utilizing a Nextflow plugin works as shown in Figure \ref{fig:Extension}. This method is similar to the method managing the measurement as part of the workflow (see Section \ref{Inside_Workflow}), but no changes are necessary to the workflow definition itself. Instead, the plugin handles writing the files used as a signal to the daemon for energy measurement. This is easier than implementing it as a direct part of the workflow, since Nextflow extensions natively support functions to execute methods before starting the workflow or individual workflow tasks, and after they have stopped running. This also omits the need to schedule the methods as additional tasks of the workflow.

\subsubsection*{Advantages and Drawbacks}

\begin{itemize}
    \item \textbf{No additional software:} The plugin is required as additional software. However, that is barely an issue since the plugin can be integrated using a single command-line parameter, and Nextflow automatically downloads and installs the plugin before executing the workflow, if the plugin has been published in the Nextflow plugins repository\footnote{https://github.com/nextflow-io/plugins, last accessed: \date{May 14, 2025}}.
    \item \textbf{Self-contained:} The approach does not rely on additional external hard- or software, making it self-contained. The only exception is the initial download of the plugin from an external source. However, the plugin can also be installed manually, removing this dependency.
    \item \textbf{Dealing with workflow faults:} The plugin can actively react to the current state of the workflow and take appropriate action if a fault occurs.
    \item \textbf{Enables full measurement:} The implementation through a Nextflow plugin technically does not allow measuring the energy consumption of the whole workflow execution, because the workflow is already running when the measurement is started. But the plugin allows starting the measurement before the first task is executed and to end it after the last task has stopped running, enabling full measurement of every single task of the workflow.
    \item \textbf{Easy portability:} Since the plugin can be integrated using a single line of code, integrating it into different workflows is trivial.
    \item \textbf{No workflow modifications:} Technically, the workflow needs to be modified to include the plugin. But since this modification consists of either a single line of code or a single command line parameter, the effort is negligible.
    \item \textbf{Low overhead:}The plugin only takes action at the beginning and end of workflow tasks, making the overhead negligible.
    \item \textbf{Multi-tenancy:} Since RAPL does not support energy measurements for individual CPU cores, an additional heuristic is necessary to assign a specific amount of energy to a physical task, if multiple tasks are running in parallel on the same CPU. The same is still true if the physical tasks belong to multiple individual workflows running in parallel.
\end{itemize}

\subsubsection{Using Prometheus}\label{Prometheus}
Prometheus\footnote{https://prometheus.io/, last accessed: \date{May 14, 2025}} is an open-source monitoring solution that automatically records time-series data of various metrics of a system and allows the user to access this data through queries. Prometheus can be deployed on each node of a compute cluster and supports reading data from Intel RAPL registers with additional configuration. By default, it can read a different set of energy values collected at the power supply. This data includes the energy consumed by the whole system. It is read through the Intelligent Platform Management Interface (IPMI) utilizing the Advanced Configuration and Power Interface (ACPI). That makes using Prometheus a fourth method to collect energy data by letting Prometheus periodically collect the data from the registers while a workflow is running. After the workflow has finished, the collected data can be accessed by connecting to the Prometheus UI and writing a query to get the energy consumption over a specific period of time on a specific node, e.g., the duration and end time of a task that was executed as part of a workflow on the node. A query for Kubernetes can look as follows:
\begin{verbatim}
sum_over_time(node_hwmon_power_average_watt{instance=~"10.0.0.37:9100|
10.0.0.38:9100"}[1476s] @ 1743495765) * 30
\end{verbatim}
This query returns the cumulative consumed energy in Joule of the nodes with the IP-addresses 10.0.0.37 and 10.0.0.38 over a period of 1476 seconds (the duration of the workflow) at the time 10:22:45 formatted as a Unix timestamp (1743495765) for a scraping interval of 30 seconds. 




\subsubsection*{Advantages and Drawbacks}

\begin{itemize}
    \item \textbf{No additional software:} In order to us Prometheus to read data regarding energy consumption from a cluster, it is necessary to install additional software on all nodes of the cluster. While this is sometimes unproblematic, it can be unwanted or even impossible in other cases, making the method utilizing Prometheus not universally applicable.
    \item \textbf{Self-contained:} Prometheus is self-contained. It runs on every node, independent of any other software. The information regarding energy consumption can be accessed directly through its own interface.
    \item \textbf{Dealing with workflow faults:} Since Prometheus is completely independent of the workflow, values from the RAPL registers are collected even if the workflow fails.
    \item \textbf{Enables full measurement:} RAPL counters are monitored continuously by Prometheus. Therefore, it is always possible to query for the period of time when the workflow or any of its tasks were executed.
    \item \textbf{Easy portability:} Prometheus works with any workflow if it is executed on the monitored cluster.
    \item \textbf{No workflow modifications:} Prometheus is completely independent of the software running on the cluster. For this reason, no modifications of the workflow are necessary.
    \item \textbf{Low overhead:} Unlike the other presented methods, Prometheus runs at all times on all nodes and continuously collects data regarding energy consumption and other metrics. This adds a significant computational overhead to the cluster, especially for high workloads \cite{monti_empirical_2023}, thereby reducing the energy efficiency of the cluster.
    \item \textbf{Multi-tenancy:} Since RAPL does not support energy measurements for individual CPU cores, an additional heuristic is necessary to assign a specific amount of energy to a physical task, if multiple tasks are running in parallel on the same CPU. The same is still true if the physical tasks belong to multiple individual workflows running in parallel.
\end{itemize}


\renewcommand{\arraystretch}{1.3} 

\begin{table}[h]
    \centering
    \caption{Strengths and Weaknesses of Energy Measurement Methods}
    \label{tab:strengths_weaknesses}
    \resizebox{\textwidth}{!}{ 
    \begin{tabular}{l m{0.4\textwidth} m{0.4\textwidth}}
        \toprule
        \textbf{Method} & \textbf{Strengths (\ding{51})} & \textbf{Weaknesses (\ding{55})} \\
        \midrule
        \textbf{As part of the Workflow} & No additional software \newline Self-contained on cluster & Not all tasks measured \newline Code integration per workflow \\
        \textbf{Wrapping Shell-Script} & No modifications required \newline Full measurement possible & Manual adjustments needed \newline Local machine communication \\
        \textbf{Nextflow Plugin} & Easy CLI integration \newline Fully measures workflow & Command pod not measured fully \\
        \textbf{Prometheus} & Continuous measurement \newline No workflow changes needed & Requires extra software \newline Increased resource consumption \\
        \bottomrule
    \end{tabular}
    }
\end{table}

\renewcommand{\arraystretch}{1.0} 

\section{Experiments}\label{Experiments}

We implemented all methods and tested them in combination with scientific workflows from three different domains of research (Genomics, Proteomics and Remote Sensing) to study the amount of energy measured by each method while executing the same workflow. We then compare the amount of measured energy between the approaches for singular tasks and the whole workflow. This enables us to draw conclusions about the viability of each method to coordinate the energy measurement and capture accurate values across all tasks of a workflow.

\subsection{Hardware Setup for Implementation}\label{Setup}
For our implementations and testing, we used a commodity cluster consisting of 14 nodes in total, of which two were available for our experiments with exclusive access. Each node of the cluster contains an Intel(R) Xeon(R) Silver 4314 CPU running at a frequency of 2.40GHz, 256GB of main memory and 8TB of mass storage. The CPUs in the cluster support the Package and DRAM domains of Intel RAPL. The cluster uses Kubernetes for orchestration, which in turn deploys Docker containers. Our workflows were programmed in Groovy using Nextflow. We deployed them to the cluster utilizing the Nextflow interface for Kubernetes (nextflow kuberun) or by executing them directly in a pod on the cluster (nextflow run), depending on the method.

\subsection{Used Workflows}\label{Used_Workflows}
To conduct our experiments, we used a set of three different scientific workflows obtained from nf-core\footnote{https://nf-co.re/pipelines/, last accessed: \date{May 14, 2025}}. The workflows are all implemented in Nextflow and can be executed on a compute cluster using Kubernetes. They feature a broad range of tasks and are used in different scientific domains. Therefore, we consider them as well suited for experiments to examine the performance of the proposed methods for energy measurement. Table \ref{tab:workflow_overview} shows an overview of the three workflows used in our experiments.

\subsubsection*{RNASeq}
RNASeq\footnote{https://nf-co.re/rnaseq/3.18.0/, last accessed: \date{May 14, 2025}} is a scientific workflow from the area of bioinformatics. It is used to analyze the RNA sequencing data obtained from organisms with a reference genome and annotation. RNASeq contains only few tasks, but some of them are very compute and memory intensive, causing long runtimes. We use a modified version of this workflow with a simplified pipeline for our tests. This pipeline only contains the tasks \textit{fastp}, \textit{star\_index}, \textit{fastqsplit}, \textit{star\_align}, \textit{samtools}, \textit{samtools\_merge} and \textit{cufflinks}. All other components contained in the nf-core version of RNASeq were removed for simplified testing.

\subsubsection*{Quantms}
Quantms\footnote{https://nf-co.re/quantms/1.2.0/, last accessed: \date{May 14, 2025}} is a bioinformatics workflow from the area of Proteomics. Among other tasks, it can be used for label-free quantification of Quantitative Mass Spectrometry data. Quantms contains a larger set of tasks than RNASeq, of which most are very small and less resource-intensive than those in RNASeq. It therefore provides a good contrast to RNASeq for our experiments.

\subsubsection*{Rangeland}
Rangeland\footnote{https://nf-co.re/rangeland/1.0.0/, last accessed: \date{May 14, 2025}} is an analysis pipeline from the area of Remote Sensing in Geography. It is used to process satellite imagery in order to assess changes in land-cover over time. During execution, the scientific workflow processes large sets of tasks in parallel, each processing one of the images in the dataset. This provides us with information about the influence of numerous small tasks, which are executed rapidly and in quick succession, on our measurement strategies.


\begin{table}[h]
    \centering
    \caption{Overview of the scientific workflows used in our experiments}
    \label{tab:workflow_overview}
    \begin{tabular}{c c c c c c}
        \toprule
        \textbf{Workflow} & \textbf{Domain} & \textbf{No. phys. Tasks} & \textbf{Input Size} & \textbf{Output Size} & \textbf{Runtime} \\
        \midrule
        RNASeq    & Bioinformatics  & 9   & 5.7~GB  & 390~MB & 25m \\
        Quantms   & Proteomics      & 61  & 100~MB  & 47~MB  & 2m 10s \\
        Rangeland & Remote Sensing  & 280 & 250~MB  & 350~MB & 3m 10s \\
        \bottomrule
    \end{tabular}
\end{table}

\subsection{Experimental Strategy}\label{Experimental_Strategy}
To compare the amount of energy measured during workflow execution by each of the four methods, we run each of the three described scientific workflows on a compute cluster. In order to keep the runtime for each test manageable (below 30 minutes per run for each workflow) and the tests realistic, we use small input datasets of real-world data. Since the four methods for energy measurement all read the same RAPL energy counters (except Prometheus), it is not necessary to run individual experiments for each method separately. Instead, we evaluate the methods simultaneously by running each workflow and collecting the values of the RAPL energy counters during execution using each of the proposed methods. This has the advantage that we save time and energy. To calculate the amount of energy captured by each of the methods (shell-script, plugin and task-based), we determine the points in time where each of the methods starts and stops the measurement. We then use the same collected RAPL values for each method to extract the exact amount of captured energy consumed by the workflow. This experiment is repeated five times for each workflow, and the averages of all runs are used during our evaluation. Note that the experiments for the method utilizing Prometheus are run separately to ensure that all data from Prometheus can be collected shortly after the experiment and using the correct scraping interval.

\subsection{Results}\label{Results}

\subsubsection*{Complete Workflow}
Our experimental results regarding the absolute and relative differences in measured energy consumption between the three approaches are shown in Table \ref{tab:energy_consumption}. The table shows that only the method based on a shell-script is able to capture the full energy consumption of the different workflows. The other two methods are missing some of the energy included in the RAPL counter values. The plugin-based and task-based methods miss about 0.19\% (754J) and 0.36\% (1437J) compared to the shell-script on RNASeq. On Quantms and Rangeland, the absolute losses are higher, with about 7.08\% (2053J) and 5.22\% (1852J) for the plugin and 7.67\% (2223J) and 5.51\% (1952J) for the task-based method, respectively. This is due to the fact that Quantms and Rangeland load additional plugins before starting the energy measurement for both the plugin-based and task-based method, delaying the start of the measurement by a few seconds (from between 3.6s and 6.2s on average for RNASeq to 9.8s and 10.8s for Quantms and 9.2s and 9.8s for Rangeland).

The higher delays causing larger differences in measured energy consumption are also part of the cause for the higher relative differences between the shell-script and the methods based on a plugin or task-based management. However, they do not explain the differences between the three workflows alone. The largest factor for the small relative difference between the approaches of only 0.36\% for RNASeq compared to 7.67\% for Quantms and 5.51\% for Rangeland is their runtime. RNASeq runs considerably longer (about 25 min) on our test dataset than Quantms (about 2 min) or Rangeland (about 3 min). Since most of the energy consumption of the tested workflows is caused by computationally intensive tasks in the middle of the workflow and not right at the start, the energy missed at the beginning of the workflow by some of the tested methods is amortized quickly, if the workflow has a sufficiently long runtime.

\begin{table}[ht]
    \centering
    \sisetup{round-mode=places, round-precision=2, table-format=6.2} 
    \caption{Energy Consumption: Absolute (J) and Relative (\%) per Workflow. Note the much lower energy consumption of Quantms and Rangeland due to the short runtime caused by very small inputs.}
    \label{tab:energy_consumption}
    \begin{tabular}{l S S S S[table-format=3.2] S[table-format=3.2] S[table-format=3.2]}
        \toprule
        \textbf{Workflow} & \multicolumn{3}{c}{\textbf{Abs Energy (J)}} & \multicolumn{3}{c}{\textbf{Rel Energy (\%)}} \\
        \cmidrule(lr){2-4} \cmidrule(lr){5-7}
        & \textbf{RNASeq} & \textbf{Quantms} & \textbf{Rangeland} & \textbf{RNASeq} & \textbf{Quantms} & \textbf{Rangeland} \\
        \midrule
        Shell-script & 393906.17 & 28981.12 & 35444.69 & 100.00 & 100.00 & 100.00 \\
        Plugin       & 393151.76 & 26928.59 & 33592.85 & 99.81  & 92.92  & 94.78  \\
        Task         & 392469.08 & 26758.27 & 33492.20 & 99.64  & 92.33  & 94.49  \\
        \bottomrule
    \end{tabular}
\end{table}

Table \ref{tab:Prometheus_results} shows the energy values extracted using Prometheus with a measurement interval of 30 seconds in comparison to those calculated with a shell-script. The amount of energy measured by Prometheus for the longer RNASeq workflow is slightly lower. This tendency is consistent across all test runs. 
When examining the two workflows with shorter runtimes, the difference between the values returned by Prometheus and the shell-script becomes much larger, with over 30\% and 36\% respectively for Quantms and Rangeland. 

\begin{table}[h]
    \centering
    \sisetup{round-mode=places, round-precision=2, table-format=6.2} 
    \caption{Comparison of measured energy consumption using Prometheus with a measurement interval of 30 seconds and a Shell-script. Note the much lower energy consumption of Quantms and Rangeland due to the short runtime caused by very small inputs.}
    \label{tab:Prometheus_results}
    \begin{tabular}{lSSS}
        \toprule
        \textbf{Dataset} & \textbf{Prometheus} & \textbf{Shell-script} & \textbf{Difference} \\
        \midrule
        RNASeq      & 380812.00J & 395829.96J & -3.94\% \\
        Quantms     & 44346.00J & 31015.71J  & 30.06\%  \\
        Rangeland   & 54432.00J & 34666.70J  & 36.31\% \\
        \bottomrule
    \end{tabular}
\end{table}

To examine how these results are affected by the measurement interval of Prometheus, we ran the same experiments with a configured measurement interval of 10 seconds. The results of these experiments are shown in Table \ref{tab:Prometheus_intervals}. Changing the measurement interval of Prometheus to 10 seconds does not reduce the difference to the energy measured using a shell-script for all workflows. While the gap between measurements becomes smaller for Quantms and Rangeland, it increases for RNASeq. This shows that the differences in energy consumption measured using Prometheus in comparison to the other methods can not be attributed to the measurement interval alone.

\begin{table}[h]
    \centering
    \sisetup{round-mode=places, round-precision=2, table-format=6.2} 
    \caption{Comparison of measured energy consumption using Prometheus with a measurement interval of 10 seconds and 30 seconds.}
    \label{tab:Prometheus_intervals}
    \begin{tabular}{lSSS}
        \toprule
        \textbf{Dataset} & \textbf{Shell-script} & \textbf{Prometheus (10s)} & \textbf{Prometheus (30s)} \\
        \midrule
        RNASeq      & 396721.28J & 376000J & 380812J \\
        Quantms     & 30115.31J & 42486.67J  & 44346J  \\
        Rangeland   & 37508.91J & 49860J  & 54432J \\
        \bottomrule
    \end{tabular}
\end{table}

For all three workflows, a shorter measurement interval leads to a reduction in measured energy consumption. This seems counterintuitive, since Prometheus saves the average energy consumption during the measurement interval at the end of each interval. The total energy consumption is then calculated by adding all saved energy values whose time stamps are during the period of measurement. Since any additional energy consumption included at the start of the measured period due to the first included interval being partly outside of the measured period should be offset at the end of the measured period, a change in the measurement interval should not lead to a consistent reduction in measured energy consumption. We hypothesize that there might be two possible reasons for the differences. The first possible reason for the difference is the method to calculate the average energy consumption during each measurement interval. At the time of writing, it is not entirely clear to the authors how these averages are calculated. It is possible that shorter measurement intervals lead to systematically smaller calculated averages for each interval. A second possibility are changes in the computational load of the machines between experiments. Since Prometheus can only collect values with one measurement interval at a time, our experiments had to be conducted one after the other, with changes to the configuration of Prometheus in between. Although we aimed to guarantee equal conditions during the tests, these changes or other external factors might have lead to a reduced base-load on the machines under test during the second part of our experiments, leading to lower energy consumption being recorded for the shorter measurement interval.

\subsubsection*{Task-based}
Utilizing the presented methods for RAPL-based energy measurement, task-based measurement is only possible for sufficiently long, non-overlapping tasks without introducing errors or resorting to heuristics. For workflows like RNASeq, where each task runs for longer than one second and each node only executes a single task at a time, the energy consumption of each task can be calculated by calculating the energy consumption on its respective node between the start time and the end time of the task. Adding up the energy consumption of all tasks results in the energy consumption for the whole workflow.

If the workflow runs tasks concurrently, their exact energy consumption can not be measured individually. Instead, it is only possible to approximate the energy consumption of a task by using heuristics to estimate how much of the total energy consumption of the node in that time window was caused by this particular task. Very short tasks like the computations on individual pictures appearing in Rangeland can cause a similar issue. If the total runtime of the task is below one second, Nextflow logs the same value for the start- and end-time of the task. In that case, it is not possible to calculate the energy consumption based on the exact runtime, but only on an estimated runtime below one second.

\section{Discussion}\label{Discussion}

\subsubsection*{Accuracy of Methods}
\begin{figure}
\centering
\includegraphics[width=1.0\linewidth, trim={0cm 0cm 0cm 0cm},clip]{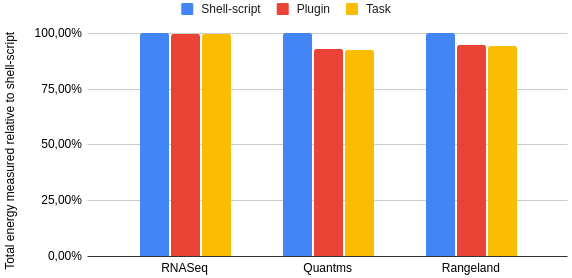}
\caption{The bar chart shows the differences in the total amount of energy measured using a plugin or an additional workflow task in relation to the amount measured using a shell-script. All values are provided as a percentage of the energy measured with a shell-script, since this method allows starting the measurement before the workflow is initialized. This enables capturing the full energy consumption of the workflow that is measurable with RAPL.}
\label{fig:Energy_Percent}
\end{figure}

In Section \ref{Measurement_Strategies} we present four ways to measure the energy consumption of the system by reading the RAPL energy counters during workflow execution on a compute cluster orchestrated by Kubernetes. Our experiments show that the measured energy consumptions of the methods using a shell-script, a plugin and task-based management are very similar. As presented in Figure \ref{fig:Energy_Percent}, even in an unfavorable situation with a very short workflow (like Quantms in our experiments), the task-based method captures more than 92\% of the energy used by the workflow. For workflows and datasets with longer runtime like our version of RNASeq, the difference in measured energy is only 0.36\%. We expect this number to become even smaller for workflows with longer runtimes, which often appear in real settings of workflow usage, since the difference is entirely caused by the delay before starting the energy measurement for some of the presented methods. This makes all three methods viable candidates for accurate energy measurement for longer workflows. However, if certainty is required that the energy measurement captures 100\% of the energy consumption that is included in RAPL energy counters during the whole workflow execution, only the method based on a shell-script can guarantee full coverage.


\begin{figure}
\centering
\includegraphics[width=1.0\linewidth, trim={0cm 0cm 0cm 0cm},clip]{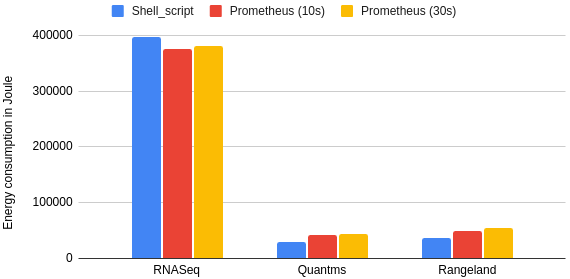}
\caption{The bar chart shows the differences in the total amount of energy measured using Prometheus with a measurement interval of 10 seconds and 30 seconds, compared to the amount measured using a shell-script.}
\label{fig:Prometheus_10s_energy}
\end{figure}

Figure \ref{fig:Prometheus_10s_energy} shows the energy consumption reported by Prometheus in comparison to that measured by the method using a shell-script, the most accurate of the proposed methods. We configured a polling interval of 30 seconds in Prometheus for our experiments. The results show that the energy consumption reported by Prometheus closely matches that of the shell-script for longer workflows such as RNASeq. It is consistently slightly lower across all runs. For short workflows such as Quantms and Rangeland, however, the energy consumption reported by Prometheus is significantly higher than that measured with the shell-script. For both workflows, there were multiple runs where the reported energy consumption was more than 40\% higher than that measured by the shell-script. We hypothesize that one contributing factor to the increased reported energy consumption is the long polling interval of 30 seconds. Since each data point reported by Prometheus represents the energy consumption of the previous 30 seconds and all data points collected during workflow execution are included when calculating the energy consumption of the workflow, some energy might be included in the data points that was consumed outside the time of workflow execution. 


To test if this energy significantly impacts results, we ran an additional set of experiments with a polling interval of 10 seconds configured in Prometheus. The energy measured in these experiments (shown in Figure \ref{fig:Prometheus_10s_energy}) is slightly lower across all workflows. This supports the hypothesis that the long polling interval of 30 seconds does affect the measured energy. However, the differences in measured energy between Prometheus and the shell-script do not decrease by much. In fact, the gap becomes larger for RNASeq. Therefore, the different results between the shell-script and Prometheus are not caused by the measurement interval, but by other differences in the measurement. Finding the exact nature of these differences remains future work.

Initial tests for the measurement of long-term and idle energy consumption show a similar pattern. The energy measured by Prometheus with a polling interval of ten seconds during idle over a period of one hour is around 36.5\% higher than that measured with RAPL. These results are consistent over multiple runs. Since we assume that Prometheus uses the same RAPL energy counters for energy measurement, the results suggest that Prometheus either uses some additional metric to include the energy consumption of other components in its output, or some unknown factor leads to higher estimations for the used energy in scenarios with low loads on the CPU and RAM. These findings underscore the need for a deeper investigation into Prometheus’s energy reporting methods, in order to ensure that the reported values are interpreted correctly.

\subsubsection*{Complexity of Implementation}
As a measure of the complexity of implementing the three methods not using dedicated cluster monitoring software, we counted the lines of code (LoC) required to implement each of them. In these numbers, we do not include the Python program to calculate the consumed energy from the RAPL energy values and the information about the workflow execution collected by Nextflow. This Python program has 381 LoC in total. Additionally, the bash-scripts used to read RAPL energy values with 24 LoC are not included either.

To implement task-based management, the amount of code required depends on the specific goal. It does require only 20 LoC in the workflow to measure from the start to the end of the workflow execution. However, it should be noted that the complexity of this implementation increases if singular tasks should be measured, due to the necessity to implement the starting and stopping procedure for each individual task. That means that task-based management requires 20 LoC for each individual task to be measured. This can result in high complexity of implementation if the workflow contains a large number of individual tasks. Also, this method must be implemented in each workflow individually.

The highest number of total lines of code are required to implement a Nextflow plugin. This method requires 105 LoC (excluding comments and empty lines) distributed over multiple files. While the structure of the Nextflow plugin is more complex than an implementation directly inside the workflow, it should be noted that most of the code can be adapted from existing plugins. This reduces the complexity of implementing a Nextflow plugin. If only the lines that need to be changed from existing plugins by the developer are counted, a Nextflow plugin is easier to implement than task-based management. In addition, a plugin has the advantage that it can be ported to a different workflow with only a single line of code in the workflow itself and no changes to the plugin. This makes a plugin the easier to implement choice if used across multiple workflows.

The method utilizing a shell-script contains 64 LoC in total. This is more than the other two methods require at the minimum, but the script contains additional code to start and monitor the workflow, delete old data, copy the files produced by the workflow and the energy measurement to the local machine and start the Python program to convert the RAPL values to meaningful energy values. This makes the script more laborious to implement initially, but it can save a lot of time in the long run by automating processes. Furthermore, it can be adapted to new workflows by changing only a single line of code.

\section{Future Work}\label{Future_Work}
In this section, we present directions for future work in order to further improve the energy measurements using RAPL and to measure and improve the energy efficiency of workflows executed on commodity clusters in general.


\subsection{Validate Accuracy of RAPL for Energy Measurement}\label{Validate_Accuracy}
One area of future work is to validate the accuracy of energy measurements using RAPL in the context of compute clusters. While the general accuracy of RAPL has been validated \cite{hackenberg_energy_2015}, it is currently unclear if RAPL provides a complete picture of the energy consumed by scientific workloads executed on compute clusters, where aspects such as the energy used by storage or the network between nodes may have a larger impact than in other environments. To validate the accuracy of RAPL, experiments comparing the energy usage measured by RAPL and hardware energy measurement tools \cite{jay_experimental_2023}, while executing scientific workflows on compute clusters, could be conducted.
While the accuracy of hardware energy measurement tools \cite{jay_experimental_2023}, RAPL \cite{david_rapl_2010, hackenberg_energy_2015} and software models \cite{lin_taxonomy_2021} has been compared in other contexts \cite{castor_estimating_2024}, there have been no such experiments in the context of workflows and cluster computing. Due to the fact that some scientific workflows are characterized by huge datasets that require much more loading of data from memory and storage as many other workloads, the absolute and relative accuracy of the methods for energy measurement might be different for these computations.


\subsection{Implement Concurrent Task-level Energy Measurement}\label{Task-level}
While our implementation is capable of measuring the energy consumption of whole workflows and single tasks that run isolated from other tasks, additional work is necessary to approximate the energy consumption of tasks that run concurrently on the same node. Since RAPL counters in the Package domain can only measure the energy of the whole CPU, it is not possible to measure these tasks separately. Instead, a heuristic is necessary to estimate for how much of the consumed energy each of the tasks is responsible. Such a metric could be based on different statistics \cite{lin_taxonomy_2021}, e.g. CPU time, utilized DRAM or the amount of CPU cores used.

\subsection{Predict Energy Consumption of Workflows}\label{Predict_Consumption}
When methods to measure the energy consumption of workflows executed on commodity clusters have been developed and proven to be reliable, the results of these measurements can be used together with metadata of the workflow executions in order to predict the energy consumption of future workflow executions based on the configuration of the workflow and the cluster. If reliable predictions of the energy consumption of a workflow are possible, they could be used to optimize the energy consumption of a workflow execution without having to execute the workflow beforehand. If runtime predictions are also possible, a two-dimensional optimization space is created that can be used to find the optimal trade-off between runtime and energy consumption. Similar work has already been published for cloud environments \cite{li_cost_2018} and utilizing dynamic voltage and frequency scaling (DVFS) \cite{pietri_energy-aware_2014}.

\section{Conclusion}\label{Conclusion}
In this work, we present four ways to read the RAPL energy counters and IPMI energy values on a compute cluster orchestrated by Kubernetes while executing a scientific workflow, in order to calculate the energy consumed by the cluster's CPU and DRAM during the execution. We discuss the workings, advantages and drawbacks of each of these approaches, experimentally show that they produce similar results and compare their complexity. We conclude that utilizing a Nextflow plugin or Prometheus offer the most benefits, if the necessary software is available. In cases where no additional software can be used, a shell-script provides a better experience to the workflow developer than implementing the energy measurement as part of the workflow. We also provide an overview of the Intel RAPL feature and highlight topics and ideas to be investigated in the future. Overall, this work helps the reader get an overview of the pitfalls of energy measurement, especially in the context of compute clusters orchestrated by Kubernetes, and provides a starting point to implement energy monitoring for workflow optimization in similar contexts.

\section*{Acknowledgements}
We would like to thank Vasilis Bountris for his valuable support in setting up the infrastructure for RAPL measurements, as well as for his technical assistance throughout the project.

Funded by the Deutsche Forschungsgemeinschaft (DFG, German Research Foundation) – Project-ID 414984028 – SFB 1404 FONDA.

\bibliographystyle{plain}

\section*{Appendix}\label{Appendix}

\begin{figure}[H]
\centering
\includegraphics[width=1.0\linewidth, trim={0cm 0cm 0cm 0cm},clip]{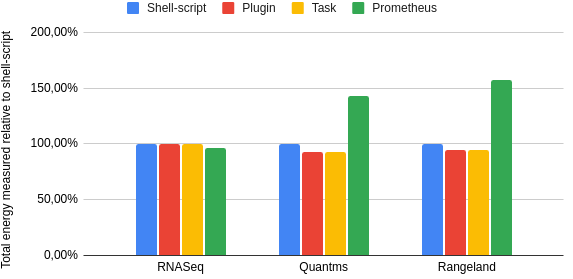}
\caption{The bar chart shows the differences in the total amount of energy measured using a plugin, an additional workflow task and Prometheus in relation to the amount measured using a shell-script. All values are provided as a percentage of the energy measured with a shell-script, since this method allows starting the measurement before the workflow is initialized. This enables capturing the full energy consumption of the workflow that is measurable with RAPL.}
\label{fig:Energy_Percent_full}
\end{figure}

\begin{figure}[H]
\centering
\includegraphics[width=1.0\linewidth, trim={0cm 0cm 0cm 0cm},clip]{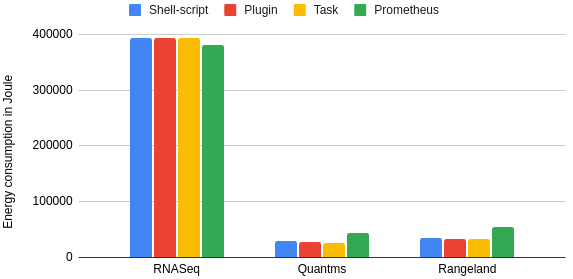}
\caption{The bar chart shows the total amount of energy measured using the four approaches on a set of three real-world workflows. The numbers presented are averages over five individual runs for each approach on each workflow.}
\label{fig:Energy_full}
\end{figure}

\end{document}